# Monte Carlo Simulation of Universal Short-Time Behavior in Critical Relaxation

Z.-B. Li[*,1], U. Ritschel[†], and B. Zheng[*]

[*]*Fachbereich Physik, Universität GH Siegen, D-57068 Siegen*

[†]*Fachbereich Physik, Universität GH Essen, D-45117 Essen*

**Abstract:** The time evolution of the three-dimensional critical Ising model relaxing from a non-equilibrium initial state is studied by means of Monte Carlo simulation. We observe the characteristic *initial increase* of the (spatially) averaged magnetization predicted by Janssen et al. [1]. The exponent $\theta'$ that governs the initial behavior is determined, and the dependence of the long-time linear decay on the initial magnetization, $m_0$, is analyzed. Our simulation corroborates earlier results derived from continuum models.

[1]On leave of absence from Zhongshan University, 510275 Guangzhou, P.R. China

Five years ago Janssen et al. [1] discovered that under certain circumstances relaxation processes in Ising-like systems near criticality show strong anomalies. When for example a system belonging to the dynamic universality class of model A [2] is quenched from an state with temperature $T \gg T_c$ to the critical point, an initially existing small magnetization will grow for a *macroscopic* time span before the actually expected decay towards equilibrium takes over. The *initial increase* of order, which was later found also in other dynamic models [3], is described by a universal power law with a new dynamic exponent, $\theta'$, that can *not* be expressed in terms of a scaling relation between the static exponents and the dynamic exponent $z$. The reason for the short-time anomaly has to be sought in the initial conditions, more precisely, in the behavior of the initial magnetization $m_0$ under renormalization-group transformations. As found by Janssen et al., the operator $m_0$ has a scaling dimension $x_0$ that is in general different from the dimension of the equilibrium magnetization, $x_\phi = \beta/\nu$. As a detailed scaling analysis reveals [4], this gives rise to a *macroscopic* time scale

$$t_0 \sim m_0^{-z/x_0} , \qquad (1)$$

and the exponent $\theta'$ is related to $x_0$ by

$$\theta' = \frac{x_0 - x_\phi}{z} . \qquad (2)$$

Numerical values for the new exponents have been obtained by means of the $\epsilon$-expansion [1].

More recently, Diehl and Ritschel [4] extended the above-mentioned considerations to systems of finite size. In the following these results which have been obtained by finite size scaling analysis and with the help of continuum field theory will be summarized because they are essential for the interpretation of the Monte Carlo data. The basic quantities that enter the scaling analysis are the time $t$, the lattice size $L$ in units of the lattice constant $a$, and the spatially averaged magnetizations $m_0$ and $m(t)$. At bulk criticality and for $L$ and $t$ much larger than any microscopic scales, one expects the following scaling form of the magnetization:

$$m(t, L, m_0) \sim l^{-\beta/\nu} \, m(l^{-z} \, t, \, l^{-1} L, \, l^{x_0} \, m_0) . \qquad (3)$$

By choosing the arbitrary rescaling parameter $l \sim t^{1/z}$, one can rewrite (3) as

$$m(t, L, m_0) \sim t^{-\beta/\nu z} \, \mathcal{F}(t/t_L, t/t_0) , \qquad (4)$$

where the initial time scale $t_0$ is given in (1), and $t_L$ is the well-known finite-size relaxation time $t_L \sim L^z$ [6].

Thus, as stated in (4), the relaxational behavior of the critical Ising system starting from a non-equilibrium initial state with small correlation length $\xi \ll L$ and initial magnetization $m_0$ is governed by *two* macroscopic time scales. This gives rise to clearly distinguishable stages of the relaxational process. First, the characteristic initial increase of order,

$$m(t) \sim t^{\theta'} , \qquad (5)$$

discovered in [1] for the infinite system, is also found in the finite system [4] when $t < t_0, t_L$. The only condition for the increase is that $t_0$ remains macroscopic which, in turn, means that $m_0$ has to be small (*cf.* (1)). Second, at time $t_{max}$ the magnetization reaches its maximum and then starts to decrease towards the equilibrium value $m(t \to \infty) = 0$. For $t_L \gg t_0$, the time $t_{max}$ is roughly given by $t_0$, but $t_{max}$ decreases with decreasing $t_L$ and can be much smaller than $t_0$ for $t_L \ll t_0$. Third, the long-time behavior for $t \gg t_L$ is described by an exponential

$$m(t \gg t_L) \approx M_\infty \, \mathrm{e}^{-t/t_L} . \qquad (6)$$

While the time scale in this "linear" regime is not influenced by the initial state [5], the amplitude $M_\infty$ is a *universal* function of both $m_0$ *and* $L$. It may be written in the form [5]

$$M_\infty(L, m_0) \approx \mathrm{const.} \; L^{-\beta/\nu} \, \mathcal{G}(t_0/t_L) . \qquad (7)$$

In the limit $t_L \gg t_0$, the scaling function $\mathcal{G}$ approaches a constant. Hence, in this case the dependence on $t_0$ (and therefore on $m_0$) drops out of the long-time behavior of $m(t)$. On the other hand, for $t_L \ll t_0$ the scaling function behaves as $\mathcal{G}(w) \approx w^{-x_0/z}$. Thus, the amplitude becomes [4]

$$M_\infty(L, m_0) \approx \mathrm{const.} \; m_0 \, L^{z\theta'} , \qquad (8)$$

i.e., it is proportional to the initial magnetization $m_0$, and the exponent $\theta'$ that in the infinitely extended system only appears in the short-time regime now governs the $L$-dependence of $M_\infty$. In other words: In the finite system there is a long-time memory of the initial condition in the sense that the asymptotically leading term depends on it.

For the Ising model the scaling function $\mathcal{G}$ has not been calculated so far. However, for the general $n$-vector model an exact solution for $n \to \infty$ has been obtained by Diehl and

Ritschel [4]. For the special case of fixed $L$, their result for the amplitude takes the simple form

$$M_\infty^{(n\to\infty)} \approx \frac{A\,m_0}{\sqrt{1 + B\,m_0^2}} \qquad (9)$$

with non-universal constants $A$ and $B$. For the Ising system ($n = 1$) we expect a qualitatively similar behavior of the asymptotic amplitude, *i.e.*, the linear dependence for small $m_0$ and asymptotically independence of the initial magnetization for large $m_0$.

Now, the question asked in this Letter is: Do the results on universal short-time behavior reported above stand up to the Monte Carlo "experiment"?

Much numerical work has been devoted to the simulation of time-displaced correlation functions *in equilibrium* [8, 9], but to our knowledge the process described above has not been studied in the literature so far. However, projects similar to ours are at present under investigation [10]. For the time being, our major aims may be summarized as follows:

- Simulation of the short-time behavior of the relaxational process starting from the non-equilibrium initial state as explained above and measurement of the exponent $\theta'$.

- Measurement of the universal amplitude of the exponential decay for late times and comparison with (9).

- Test of the scaling form (3) and thereby determine the scaling dimension $x_0$.

An initial state with non-vanishing magnetization and small correlation length in principle can be generated by letting evolve the system to thermal equilibrium with temperature $T \gg T_c$ within an external magnetic field. However, this procedure turns out to be unpractical because it is too time consuming, and, more importantly, due to the finite number of degrees of freedom, even for $T \gg T_c$ and strong magnetic field, fluctuations of the initial magnetization of the canonical ensemble can not be neglected. As a consequence, the width of the distribution of the initial magnetization gives rise to an additional scale which may alter the simple relation (3). (In the continuous system the same phenomenon occurs if the initial temperature is close to $T_c$ [1, 4].) In order to avoid this, we start from a "microcanonical" initial state with *fixed* $m_0$ and small correlation length. An initial configuration with a given $m_0$ is generated as follows. First, the spins on single lattice sites are determined independently, with probability $1/2 + m_0/2$ for $S_i = 1$ and $1/2 - m_0/2$ for $S_i = -1$. Afterwards, if the avarage magnetization is below or above the desired value $m_0$, single spins are chosen randomly and flipped if $S_i = -1$ or $S_i = +1$, respectively, until the average magnetization is exactly $m_0$.

Then the standard heat-bath formalism is used to trace the time evolution of the system [7]. Time is measured as usual in Monte Carlo steps per spin (MCS) and measurements are carried out after each sweep. We work at the (bulk) critical temperature $J/T_c = 0.2216$ [11]. For each magnetization profile we have averaged over 150,000 to 400,000 statistically independent histories of the system or runs, up to $t = 100 - 200$ MCS. Further, for each run a new initial configuration is generated and a new sequence of random numbers is used for the ensuing relaxational process.

Profiles for various $L$ and $m_0$ are displayed in Fig. 1 and 2. In Fig. 1 both axis are logarithmic to show clearly the power-law behavior (5) for early times, whereas in Fig. 2 only the $m$-axis is plotted logarithmically to bring out the linear decay for long times. All curves show the initial increase, and $\theta'$ is determined by fitting a straight line to the short-time regime.

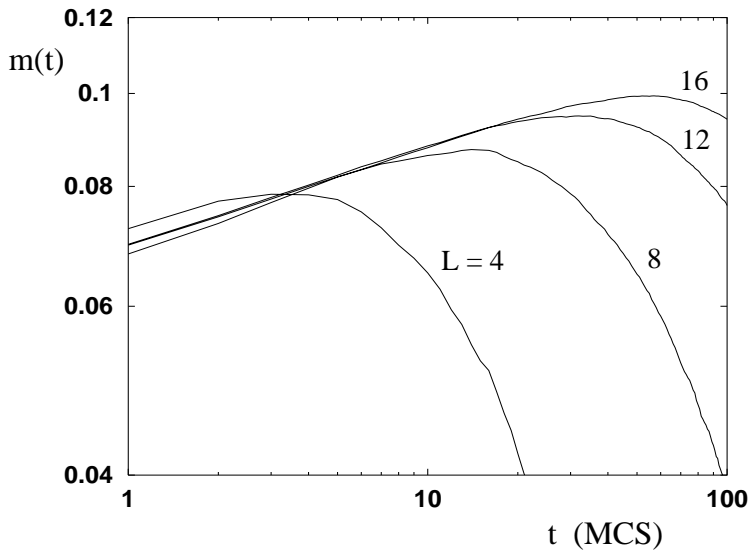

**Fig. 1:** Magnetization profiles for $m_0 = 0.06$ and $L = 4, 8, 12, 16$ in double-logarithmic representation.

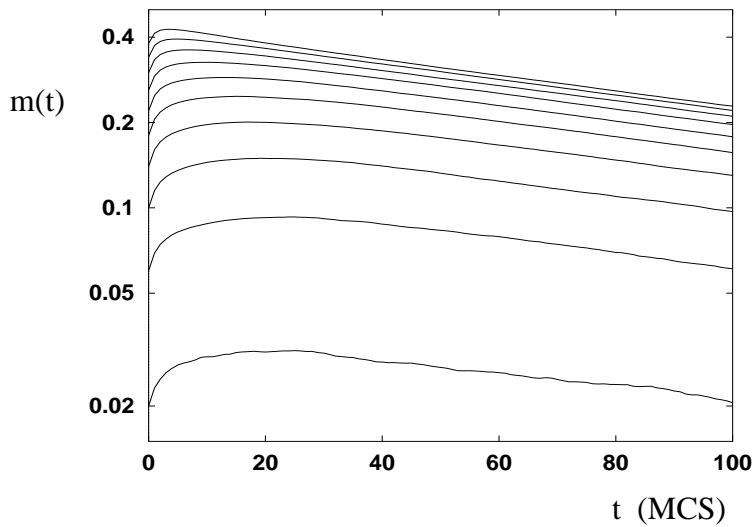

**Fig. 2:** Magnetization profiles for $L = 10$ and $m_0 = 0.02, 0.06, \ldots, 0.38$ in semi-logarithmic representation.

Because we expect the universal initial increase only for small $m_0$, we calculate $\theta'$ from eight profiles with $0.02 \leq m_0 \leq 0.10$ and $4 \leq L \leq 16$. The results and their statistical errors are displayed in Table 1. (The figure for $m_0 = 0.04$ and $L = 10$ is included in the table but the corresponding curve is not shown in Fig. 2.) They have to be compared with 0.08 from first-order and 0.13 from second-order $\epsilon$-expansion [1]. In the range of $L$ and $m_0$ we consider, the scattering of the data is small. Closest to the truth is probably the $\theta' = 0.102(2)$ from $L = 16$ and $m_0 = 0.06$.

| $L = 10$ | | $m_0 = 0.06$ | | Scaling Analysis | | |
|---|---|---|---|---|---|---|
| $m_0$ | $\theta'$ | $L$ | $\theta'$ | $(L, L')$ | $x_0$ | $\theta'$ |
| 0.02 | 0.115(11) | 16 | 0.102(2) | (16, 6) | 0.725(13) | 0.100(6) |
| 0.04 | 0.105(5) | 12 | 0.102(3) | (10, 6) | 0.738(26) | 0.107(13) |
| 0.06 | 0.106(5) | 8 | 0.114(10) | (16, 10) | 0.710(22) | 0.093(11) |
| 0.10 | 0.102(6) | 4 | 0.096(43) | | | |

**Table 1:** Monte Carlo results for $\theta'$ and $x_0$.

The dependence of the amplitude of the linear decay has been obtained from the data of Fig. 2. The results which are displayed in Fig. 3 clearly show the linear increase for small $m_0$ and the tendency to become independent of $m_0$ for larger values. Fitted to the numerical data is the analytic large-$n$ solution (9) with constants $A = 1.98$ and $B = 14.2$. We actually do not expect that the large-$n$ result also holds in exactly the same form for the Ising system, but apparently the scaling function for the latter is quite similar.

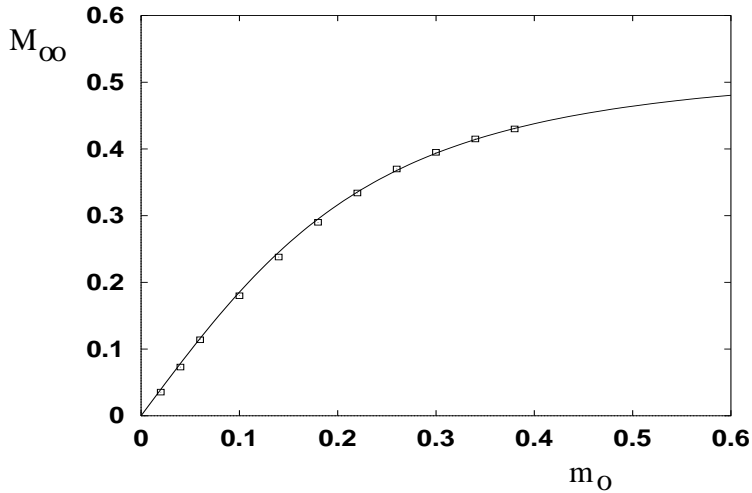

**Fig. 3:** Amplitude of the linear decay as a function of $m_0$ for $L = 10$. The squares respresent the Monte Carlo results. The function fitted to them is the exact large-$n$ solution (9).

Finally, we test the scaling form (3) and determine directly the scaling dimension $x_0$ by comparing profiles from systems of different size. If (3) is satisfied, it should be possible to map curves onto each other by appropriate rescalings of $m_0$, $m$ and $t$. To this end, we have calculated one profile for lattice size $L$ and initial magnetization $m_0$ and several curves for a smaller lattice $L'$ and a certain range of $m'_0$. Now the curves from the primed system can be mapped to the unprimed system by (i) multiplying by an overall factor $b^{\beta/\nu}$ and (ii) by stretching the time axis by $b^z$ with $b = L/L'$. Here we have used the literature values $z = 2.04$ [8] and $\beta/\nu = 0.52$ [11], which are, by the way, consistent with our own data. For one example the result is shown in Fig. 4. The solid line is from $L = 16$ and $m_0 = 0.06$, and the three dashed lines are those from $L' = 10$ and $m'_0 = 0.082$, 0.084 and 0.086, respectively. (Note that due to the finite number of the degrees of freedom the increment for changes of $m'_0$ here is 0.002.) After matching the magnetizations by linear interpolation of $m'_0$, $x_0$ is determined with the help of $m'_0 = (L/L')^{x_0} m_0$. The results for $x_0$ and $\theta'$ - the latter has been computed with the help of (2) - obtained from $(L, L') = (16,10)$, $(16,6)$ and $(10,6)$ are displayed in Table 1. Again this may be compared to 0.67 and 0.77 from first-order and second-order $\epsilon$-expansion.

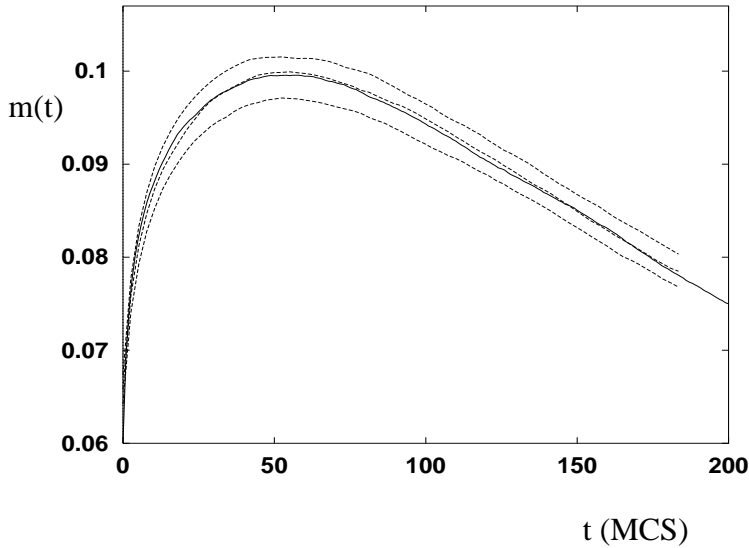

**Fig. 4:** Test of finite-size scaling relation (3). Compared are profiles for $L = 16$ and $m_0 = 0.06$ (solid line) and $L = 10$ and $m_0 = 0.082, 0.084, 0.086$ (dashed lines, from bottom to top). For details see text.

In summary, we have investigated the influence of the initial conditions on the relaxational behavior of the three-dimensional Ising model by means of Monte Carlo simulation. The measured values of the exponents $\theta'$ and $x_0$ are consistent with those found by renormalization-group improved perturbation theory [1]. As demonstrated in Fig. 3, also the dependence of the amplitude of the linear decay on the initial magnetization agrees very well with the expectation from continuum calculations [4, 5]

Compared with what is standard today in Monte Carlo simulations in critical dynamics [8, 9], the lattices we have studied are relatively small. However, we think the results derived from our small systems are very well suited as a first step to verify the anomalous initial behavior and other universal properties related to it. Especially, we point out that the dependence of $\theta'$ on the lattice size in the range of $L$ studied is very weak (*cf.* Table 1). Thus, we do not expect qualitative or dramatic quantitative changes for the simulation on larger lattices.

*Acknowledgement:* Z. B. L. would like to thank the Alexander-von-Humboldt-Stiftung for a fellowship. U. R. is grateful to the Deutsche Forschungsgemeinschaft for partial support through Sonderforschungsbereich 237.

# References

[1] Janssen H K, Schaub B and Schmittmann B 1989 *Z. Phys. B* **73** 539

[2] Hohenberg P C and Halperin B I 1977 *Rev. Mod. Phys.* **49** 435

[3] Oerding K and Janssen H K 1993 *J. Phys. A* **26** 3369; *ibid.* **26** 5295

[4] Diehl H W and Ritschel U, 1993 *J. Stat. Phys.* **73** 1

[5] Ritschel U and Diehl H W 1994 *Long-time traces of the initial condition in relaxation phenomena near criticality* Essen preprint

[6] Sancho J M, San Miguel M and Gunton J D 1980 *J. Phys. A* **13** L443;
Diehl H W 1987 *Z. Phys. B* **66** 211;
Goldschmidt Y Y 1987 *Nucl. Phys. B* **280**, 340; *ibid.* **285** 519

[7] Binder K and Heermann D W 1992 *Monte Carlo Simulation in Statistical Physics* (Berlin: Springer)

[8] Wansleben S and Landau D P 1991 *Phys. Rev. B* **43** 6006

[9] Heuer H O, 1992 *J. Phys. A* **25** L567; 1993 J. Stat. Phys. **72** 789

[10] Ruge C and Oerding K: private communications

[11] Ferrenberg A M and Landau D P 1991 *Phys. Rev. D* **44** 5081